\documentclass[superscriptaddress,twocolumn,showpacs,pra]{revtex4-1}
\usepackage[latin9]{inputenc}
\usepackage{amssymb}
\usepackage{graphicx}

\makeatletter
 
 \@ifundefined{textcolor}{}
 {%
   \definecolor{BLACK}{gray}{0}
   \definecolor{WHITE}{gray}{1}
   \definecolor{RED}{rgb}{1,0,0}
   \definecolor{GREEN}{rgb}{0,1,0}
   \definecolor{BLUE}{rgb}{0,0,1}
   \definecolor{CYAN}{cmyk}{1,0,0,0}
   \definecolor{MAGENTA}{cmyk}{0,1,0,0}
   \definecolor{YELLOW}{cmyk}{0,0,1,0}
 }

\usepackage{dcolumn}
\usepackage{bm}\usepackage{epstopdf}

\makeatother

\begin{document}

\title{Quantum heat haths satisfying the eigenstate thermalization hypothesis}

\author{O. Fialko }

\affiliation{Institute of Natural and Mathematical Sciences and Centre for Theoretical
Chemistry and Physics, Massey University, Auckland, New Zealand}
\begin{abstract}
A class of autonomous quantum heat baths satisfying the eigenstate
thermalization hypothesis (ETH) criteria is proposed. We show that
such systems are expected to cause thermal relaxation of much smaller
quantum systems coupled to one of the baths local observables. The
process of thermalization is examined through residual fluctuations
of local observables of the bath around their thermal values predicted
by ETH. It is shown that such fluctuations perturb the small quantum
system causing its decoherence to the thermal state. As an example,
we investigate theoretically and numerically thermalization of a qubit
coupled to a realistic ETH quantum heat bath. 
\end{abstract}

\pacs{03.65.Aa, 05.70.-a, 07.20.-n}

\maketitle

\section{Introduction}

Quantum thermodynamics has attracted much attention in recent years.
On the one hand, there has been much progress in understanding the
origin of thermalization in closed quantum systems \cite{Eisert15}.
Thermalization is meant to be based purely on the phenomenology of
quantum mechanics without any need for any external sources of heat
to cause thermal relaxation of local observables in a quantum system
\cite{Jensen85}. It was shown that it is the properties of eigenstates
of an isolated quantum system whose local observables show thermal
behavior \cite{Srednicki94,Tasaki98}. Later, a general canonical
principle was introduced \cite{Golstein06}, stating that local observables
thermalize for almost all pure states of a sufficiently big closed
system under a global constrain. In this situation, the whole system
with respect to local observables can be regarded to be in a microcanonical
ensemble. On the other hand, in the studies of quantum thermal machines
the emphasis is put on the properties of small systems which are coupled
to some macroscopic sources of heat \cite{Kosloff13}. The heat baths
are treated phenomenologically using the  methods of open quantum
systems. They are usually a collection of an infinite number of harmonic
oscillators, which are presupposed to be in thermal equilibrium \cite{breuer02}. 

The smallest thermal machines are supposed to work at atomic scales
and be of benefits in nanotechnology \cite{Linden10}. To be able
to do useful work they need to extract heat from heat baths. So far
the role of heat baths has been played by macroscopic objects such
as electromagnetic radiation or beams of atoms \cite{Scully03}. An
interesting question arises whether it is possible to realize isolated
quantum heat baths as sources of heat for such small machines. Combining
together such small machines and microscopic heat baths it would be
possible to create genuine quantum heat devices. A quantum heat bath
must satisfy certain conditions to be able to thermalize a small quantum
system coupled to it. Among them is the requirement that energies
of the bath must be much denser than energies of the system and its
density of states must be an increasing function of energy. Thermalization
of the small system is then achieved by choosing a bath operator that
couples to the system to be a random matrix \cite{Gemmer06}. The
combined bath-system Hamiltonian is then itself random with typically
thermal behavior \cite{Jensen85}. Indeed, it has been shown in Ref.
\cite{Wu14} that a large class of isolated quantum systems in a pure
state can serve as a heat bath for its subsystem. The crucial limitation
of these works is that the bath and the system could not be separated;
they were studied together as a closed quantum system. According to
the second law of thermodynamics it is not possible to extract work
from a single bath in a complete cycle; we need at least two baths
to be able to build a thermal machine. The need for autonomous heat
baths arises. In this paper, we present a class of finite realistic
quantum heat baths; those thermal properties are dictated solely by
the properties of their eigenstates. We show that such systems, whose
eigenstates satisfy the ETH criteria, cause thermal relaxation of
smaller systems weakly coupled to it. The ETH bath is thermal with
some well-defined temperature regardless whether a small system is
coupled to it or not. Therefore, the ETH bath is autonomous, the case
which has not been addressed previously. This opens the possibility
to use such baths to build thermal machines. ETH concerns the thermal
properties of local observables of the bath, while the qubit is a
separate system not covered by the ETH requirements. In this paper
we show when and how the ETH bath thermalizes the qubit coupled to
it.

ETH states that under certain conditions almost all eigenstates of
a closed quantum system with some Hamiltonian $\hat{H}_{B}$ have
the properties of a thermal state. Let $|E_{k}\rangle$ be an eigenstate
of such system with eigenvalue $E_{k}$ and $\hat{O}$ its local operator.
According to ETH if the diagonal elements ${\cal O}(E_{k})\equiv\langle E_{k}|\hat{O}|E_{k}\rangle$
are smooth functions of $E_{k}$, while the off-diagonal elements
$\langle E_{k}|\hat{O}|E_{l}\rangle$ are negligibly small, then for
an initial state $|\phi_{0}\rangle$ with energy $E_{B}$ and small
energy variance the expectation value $\langle\phi|\hat{O}|\phi\rangle$
in the pure state $|\phi\rangle=\exp(-it\hat{H}_{B}/\hbar)|\phi_{0}\rangle$
relaxes to a steady thermal value and stays there for almost all times
\cite{Srednicki94}. ETH predicts that the relaxed value tends to
an appropriate microcanonical ensemble average $\langle\phi|\hat{O}|\phi\rangle\approx\frac{1}{{\cal N}}\sum_{k}^{'}\langle E_{k}|\hat{O}|E_{k}\rangle.$
Here we sum over ${\cal N}$ energy eigenstates near energy $E_{B}$
of the system, $E_{k}\in[E_{B}-\delta,E_{B}+\delta]$ with fixed $\delta\ll E_{B}.$
This yields that the whole isolated system with respect to a local
observable can be regarded to be in the microcanonical ensemble with
the density matrix

\begin{equation}
\hat{\rho}_{m}=\frac{1}{{\cal N}}\sum_{k}^{'}|E_{k}\rangle\langle E_{k}|.\label{eq:MDmatrix}
\end{equation}
For a finite quantum system, the values of the matrix elements $O_{kl}\equiv\langle E_{k}|\hat{O}|E_{l}\rangle$
are not ideally smooth functions of eigenvalues $E_{k}$ but exhibit
some residual fluctuations: 

\begin{equation}
O_{kl}={\cal O}(E_{k})\delta_{kl}+{\cal R}_{kl},\label{eq:Fluctuations}
\end{equation}
where ${\cal R}_{kl}$ are random with zero mean and small according
to ETH. The semiclassical analysis shows that the values $|{\cal R}_{kl}|^{2}$
are characterized by a smooth function \cite{Feingold86}

\begin{equation}
|{\cal R}_{kl}|^{2}\simeq{\cal S}(E_{k}-E_{l})/2\pi\rho(\overline{E}),\label{eq:R}
\end{equation}
where ${\cal S}(x)$ is the spectral function (see Ref. \cite{Feingold86}
for more details), $\rho(x)$ is the density of states and $\overline{E}=(E_{k}+E_{l})/2$.
The density of states grows exponentially with the number of particles;
therefore, the fluctuations are small if the number of particles is
large. Recent developments have demonstrated that this behavior pertains
to chaotic quantum systems lacking classical analogs \cite{Santos12}.
The fluctuating behavior in Eq. (\ref{eq:Fluctuations}) manifests
itself on the expectation value of the observable

\begin{equation}
\langle\phi|\hat{O}|\phi\rangle=\sum_{k}|\alpha_{k}|^{2}{\cal O}(E_{k})+\sum_{k,l}\alpha_{k}^{\ast}\alpha_{l}e^{it(E_{k}-E_{l})/\hbar}{\cal R}_{kl}.\label{eq:Fluctuation2}
\end{equation}
The first term here is the expected relaxed thermal value, while the
second term averages to zero at long times. However, the fluctuations
$\langle\phi|\hat{O}^{2}|\phi\rangle-\langle\phi|\hat{O}|\phi\rangle^{2}\simeq\sum_{kl}|\alpha_{k}|^{2}|\alpha_{l}|^{2}|{\cal R}_{kl}|^{2}$
are not negligible. They can be interpreted as residual thermal fluctuations
\cite{Feingold86}.

\section{ETH Bath}

Let an arbitrary small quantum system with Hamiltonian $\hat{H}_{S}$
be coupled to a larger quantum system satisfying ETH. For brevity,
we call the latter ETH bath. The combined system Hamiltonian is given
by

\begin{equation}
\hat{H}=\hat{H}_{{\rm S}}+\hat{H}_{I}+\hat{H}_{{\rm B}},\label{eq:Hfull}
\end{equation}
where $\hat{H}_{I}=g\hat{X}_{{\rm S}}\otimes\hat{O}$ is the interaction
between the system and the bath. Here $g$ is an interaction strength;
$\hat{X}_{S}$ is a system operator which couples to the bath operator
$\hat{O}$. Let $|\epsilon_{l}\rangle$ be the eigenstates of the
system and $|E_{k}\rangle$ be the eigenstates of the bath. The basis
of the combined system is chosen to be $|{\cal E}_{lk}\rangle=|\epsilon_{l}\rangle\otimes|E_{k}\rangle$.
Using Eq. (\ref{eq:Fluctuations}), the combined Hamiltonian assumes
the form of the sum of a ``smooth'' part with the elements $\langle lk|\hat{H}|l'k'\rangle=\delta_{kk'}[\delta_{ll'}(\epsilon_{l}+E_{k})+gX_{ll'}{\cal O}(E_{k})]$
and the ``irregular'' part with the elements $gX_{ll'}{\cal R}_{kk'}$.
Here, we denoted $X_{ll'}=\langle\epsilon_{l}|\hat{X}_{S}|\epsilon_{l'}\rangle$. 

The smooth part is a block matrix which contains for a given index
$k$ a smaller matrix $h_{ll'}(E_{k})=\delta_{ll'}(\epsilon_{l}+E_{k})+gX_{ll'}{\cal O}(E_{k})$.
The terms $gX_{ll'}{\cal O}(E_{k})$ simply shift the energies of
the unperturbed system by an amount $\delta\epsilon_{l}(E_{k})$,
which are some smooth functions of $E_{k}$. For an ETH bath the energy
variance is small around mean energy $E_{B}$ and we can assume that
its eigenenergies $E_{k}\approx E_{B}$. Since the coupling $g$ is
small we may apply perturbation theory to estimate the energy shift

\begin{equation}
\delta\epsilon_{l}(E_{B})\approx gX_{ll}O(E_{B})+g^{2}O^{2}(E_{B})\sum_{l'}\frac{|X_{ll'}|^{2}}{\epsilon_{l}-\epsilon_{l'}}.\label{eq:Perturb}
\end{equation}

The irregular part, on the other hand, intertwines blocks with different
indices $k$ in a chaotic manner due to randomness of ${\cal R}_{ll'}$.
This has a more profound effect on the dynamical evolution of the
small quantum system. One can imagine that these random terms disturb
the phase of the quantum state of the small system leading to its
dephasing. Additionally, the off-diagonal nature of the irregular
part allows energy exchange between the bath and the small system
leading to dissipation. Both effects, dephasing and dissipation, lead
to decoherence of the small system \cite{Schlosshauer13}. We will
show that the ETH bath causes the system to decohere to a thermal
state.

\section{Qubit coupled to ETH bath}

Consider a specific example of great interest: a qubit coupled to
the above ETH quantum bath. The Hamiltonian of the qubit with two
available states $|1\rangle$ and $|2\rangle$ and corresponding energies
$\epsilon_{1}$ and $\epsilon_{2}$ reads

\begin{equation}
\hat{H}_{{\rm S}}=\epsilon_{1}|1\rangle\langle1|+\epsilon_{2}|2\rangle\langle2|,
\end{equation}
where $\Delta\equiv\epsilon_{2}-\epsilon_{1}>0$. We allow energy
exchange between the qubit and the bath considering for simplicity
$\hat{X}_{S}=\sigma_{+}+\sigma_{-}$, where $\sigma_{+}=|2\rangle\langle1|$
and $\sigma_{-}=|1\rangle\langle2|$ are the raising and lowering
operators of the qubit. Following the formalism of open quantum systems
\cite{breuer02}, we move into the interaction picture $\hat{H}_{I}(t)=e^{it(\hat{H}_{S}+\hat{H}_{B})/\hbar}\hat{H}_{I}e^{-it(\hat{H}_{S}+\hat{H}_{B})/\hbar}$,
which yields $\hat{X}_{S}(t)=e^{i\Delta t/\hbar}\sigma^{+}+e^{-i\Delta t/\hbar}\sigma^{-}$
and $\hat{O}(t)=\sum_{k,l}e^{it(E_{k}-E_{l})/\hbar}O_{kl}|k\rangle\langle l|$.
As the next step, we need to calculate the correlation function of
the bath operators $C(t,s)={\rm tr}[\hat{O}(t)\hat{O}(s)\hat{\rho}_{B}]$.
As we discussed above, the bath satisfying ETH can be assumed to be
in the microcanonical ensemble $\hat{\rho}_{B}=\hat{\rho}_{m}$, where
the density matrix $\hat{\rho}_{m}$is given in Eq. (\ref{eq:MDmatrix}).
This allows us to evaluate the correlation function

\begin{equation}
C(t,s)=C(t-s)=\frac{1}{{\cal N}}\sum_{k}^{'}\sum_{l}e^{i(E_{k}-E_{l})(t-s)/\hbar}|O_{kl}|^{2}.
\end{equation}

The coupling is assumed to be small and the bath is large enough to
satisfy ETH requirements. Under these conditions we use the Born-Markov
approximation to study the dynamics \cite{breuer02}. In fact, it
was shown in Ref. \cite{Gemmer08} that randomness of the coupling
renders the second order approach valid. Under these assumptions we
obtain the following Schr\"odinger picture master equation 
\begin{figure}
\includegraphics[width=0.9\columnwidth]{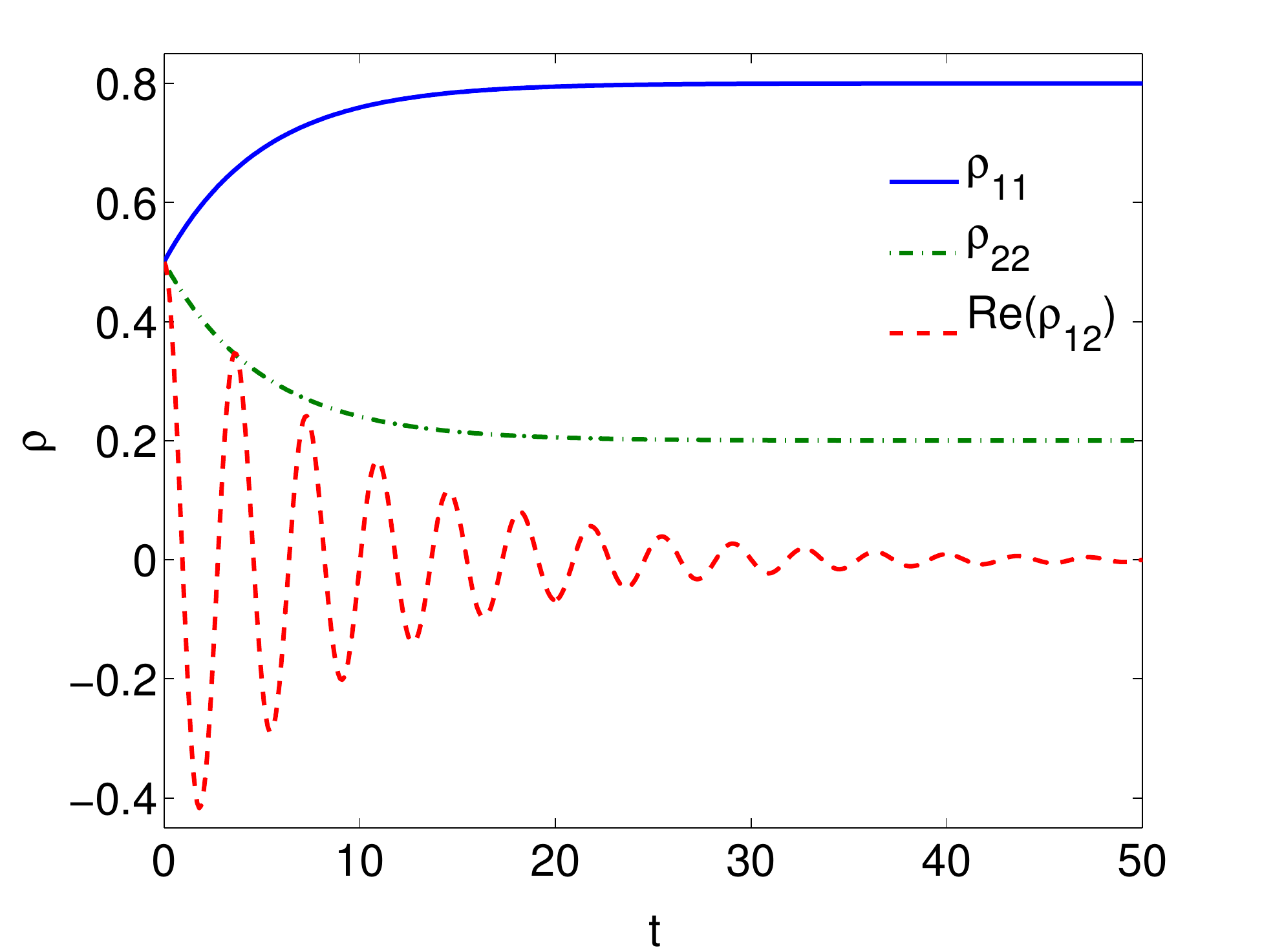}

\protect\caption{Elements of the density matrix of the qubit, which were obtained by
solving Eq. (\ref{eq:MasterRHODiag}) and (\ref{eq:MasterRHOOFFDiag})
with the initial state $\rho_{11}=\rho_{22}=\rho_{12}=\rho_{21}=0.5$
and $\exp(\beta\Delta)=4$, $\gamma=0.1$, $\delta\Delta=0.5$, and
$\Delta=1$. The qubit relaxes to the thermal state $\rho_{11}/\rho_{22}=4$
and $\rho_{12}=\rho_{21}=0.$ The process of thermalization (relaxation
of diagonal elements) is twice faster than the process of decoherence
(relaxation of off-diagonal elements). \label{fig:MasterEQ}}
\end{figure}

\begin{eqnarray}
\hbar\frac{\partial}{\partial t}\hat{\rho}_{S}(t) & = & -i[\hat{H}_{S},\hat{\rho}_{S}(t)]\nonumber \\
 & - & g^{2}\int_{0}^{\infty}d\tau[\hat{X}_{S},\hat{X}_{S}(-\tau)\hat{\rho}_{S}(t)]C(\tau)\nonumber \\
 & - & g^{2}\int_{0}^{\infty}d\tau[\hat{\rho}_{S}(t)\hat{X}_{S}(-\tau),\hat{X}_{S}]C(-\tau).\label{eq:MasterEQ}
\end{eqnarray}
To solve this equation we need to evaluate the integrals $\int_{0}^{\infty}d\tau\hat{X}_{S}(-\tau)C(\pm\tau)$,
which arise in the second and third lines of Eq. (\ref{eq:MasterEQ}).
This boils down to taking the integrals $I(\Delta)=\int_{0}^{\infty}d\tau e^{\pm i\Delta\tau/\hbar}C(\tau)$.
By using the identity $\int_{0}^{\infty}dte^{\pm i\epsilon t}=\pi\delta(\epsilon)\pm i\frac{{\cal P}}{\epsilon}$,
where ${\cal P}$ denotes the principal value, we obtain

\begin{eqnarray}
I(\Delta) & = & {\cal S}(\Delta)\frac{1}{2{\cal N}}\sum_{k}^{'}\rho(E_{k}\pm\Delta)/\rho(E_{k}\pm\Delta/2)\nonumber \\
 & + & \frac{1}{{\cal N}}{\cal P}\sum_{k}^{'}\sum_{l}\frac{|O_{kl}|^{2}}{E_{k}-E_{l}\pm\Delta}.\label{eq:I}
\end{eqnarray}
Here, $\rho(\epsilon)=\sum_{l}\delta(\epsilon-E_{l})$ is the density
of states of the bath and we used Eqs. (\ref{eq:Fluctuations}) and
(\ref{eq:R}) in deriving Eq. (\ref{eq:I}). The first term in $I(\Delta)$
is equal to ${\cal S}(\Delta)/2e^{\pm\beta\Delta/2}$, where $\beta=\frac{\partial\log\rho(E_{B})}{\partial E_{B}}$
is the inverse temperature of the bath. We have assumed that $\Delta\ll E_{B}$,
which allowed us to expand $\rho(E_{k}\pm\Delta)\approx\rho(E_{k})(1\pm\beta\Delta)\approx\rho(E_{k})e^{\pm\beta\Delta}$
and similarly for $\rho(E_{k}\pm\Delta/2)\approx\rho(E_{k})e^{\pm\beta\Delta/2}$.
The second term in $I(\Delta)$ is significant if $E_{k}=E_{l}$ when
it is equal to $\pm i\frac{1}{{\cal N}}\sum_{k}^{'}O^{2}(E_{k})/\Delta\approx\pm iO^{2}(E_{B})/\Delta$.
\begin{figure}
\includegraphics[width=0.9\columnwidth]{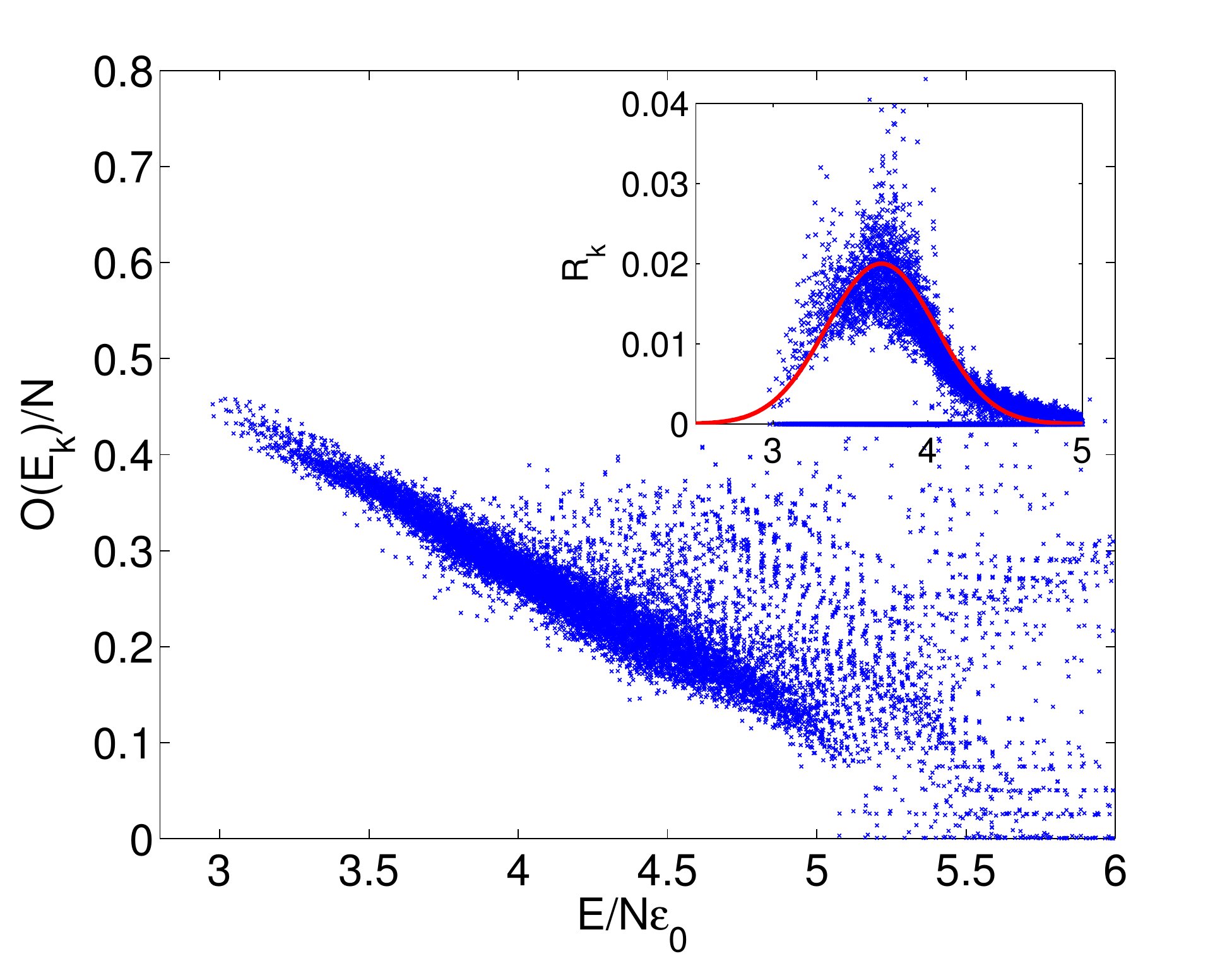}

\protect\caption{Example of expectation values of the bath operators in the eigenenergy
basis, $\langle E_{k}|\hat{b}_{L}^{0\dagger}\hat{b}_{L}^{0}|E_{l}\rangle={\cal O}(E_{k})\delta_{kl}+{\cal R}_{kl}$.
At low energies the distribution of the diagonal elements ${\cal O}(E_{k})$
resembles smooth function in accordance with ETH, while at larger
energies their behavior is more chaotic. Inset: off-diagonal elements
$R_{k}=|{\cal R}_{kk_{0}}|^{2}$ with fixed $k_{0}$ corresponding
to $E_{k_{0}}\approx3.65N\epsilon_{0}$. They are much smaller than
the diagonal elements in accordance with ETH and their distribution
resembles a smooth symmetric function around $E_{k_{0}}$ (shown in
red) in accordance with the semiclassical estimate. \label{fig:ETH}}
\end{figure}

From Eq. (\ref{eq:MasterEQ}) we can now obtain two coupled equations
for the diagonal elements of the density matrix describing the process
of thermalization: 

\begin{eqnarray}
\hbar\frac{\partial\rho_{11}}{\partial t} & = & -g^{2}{\cal S}(\Delta)e^{-\beta\Delta/2}\rho_{11}+g^{2}{\cal S}(\Delta)e^{\beta\Delta/2}\rho_{22},\nonumber \\
\hbar\frac{\partial\rho_{22}}{\partial t} & = & -g^{2}{\cal S}(\Delta)e^{\beta\Delta/2}\rho_{22}+g^{2}{\cal S}(\Delta)e^{-\beta\Delta/2}\rho_{11}.\label{eq:MasterRHODiag}
\end{eqnarray}
The qubit relaxes to the expected thermal steady state $\lim_{t\rightarrow\infty}\rho_{11}(t)/\rho_{22}(t)=e^{\beta\Delta}$.
The specific form of fluctuations $|{\cal R}_{kl}|^{2}$ used in the
derivation of this result proves to be crucial to yield the correct
thermal steady state. The remaining equations are for the off-diagonal
elements describing the process of decoherence:

\begin{eqnarray}
\hbar\frac{\partial\rho_{12}}{\partial t} & = & i(\Delta+2\delta\Delta)\rho_{12}+i2\delta\Delta\rho_{21}-\gamma(\rho_{12}-\rho_{21}),\nonumber \\
\hbar\frac{\partial\rho_{21}}{\partial t} & = & -i(\Delta+2\delta\Delta)\rho_{21}-i2\delta\Delta\rho_{12}-\gamma(\rho_{21}-\rho_{12}),\label{eq:MasterRHOOFFDiag}
\end{eqnarray}
where we denoted $\delta\Delta=g^{2}O^{2}(E_{B})/\Delta$ and $\gamma=g^{2}{\cal S}(\Delta)\cosh(\beta\Delta/2)$.
It can be shown that the population relaxation occurs at the time
scale $1/(2\gamma)$, while the loss of coherence occurs at the time
scale $1/\gamma$. Thermalization of the qubit is thus twice faster
than decoherence, which is a standard result in the context of atomic
decay and quantum optics \cite{breuer02}. An example of this dynamics
is shown in Fig. \ref{fig:MasterEQ}. The off-diagonal terms can be
shown to behave as $\sim\exp(it\sqrt{\Delta(\Delta+4\delta\Delta)-\gamma^{2}}/\hbar)\exp(-\gamma t/\hbar).$
At small $g$ we have $\gamma\ll\delta\Delta\ll\Delta$, therefore
the energy levels of the qubit are shifted by the amount $\sqrt{\Delta(\Delta+4\delta\Delta)-\gamma^{2}}-\Delta\approx2\delta\Delta\equiv2O^{2}(E_{B})/\Delta$.
This can be easily seen also from Eq. (\ref{eq:Perturb}), since in
this case $|X_{12}|=1$ and $\epsilon_{2}-\epsilon_{1}\equiv\Delta$.
\begin{figure}[t]
\includegraphics[width=0.9\columnwidth]{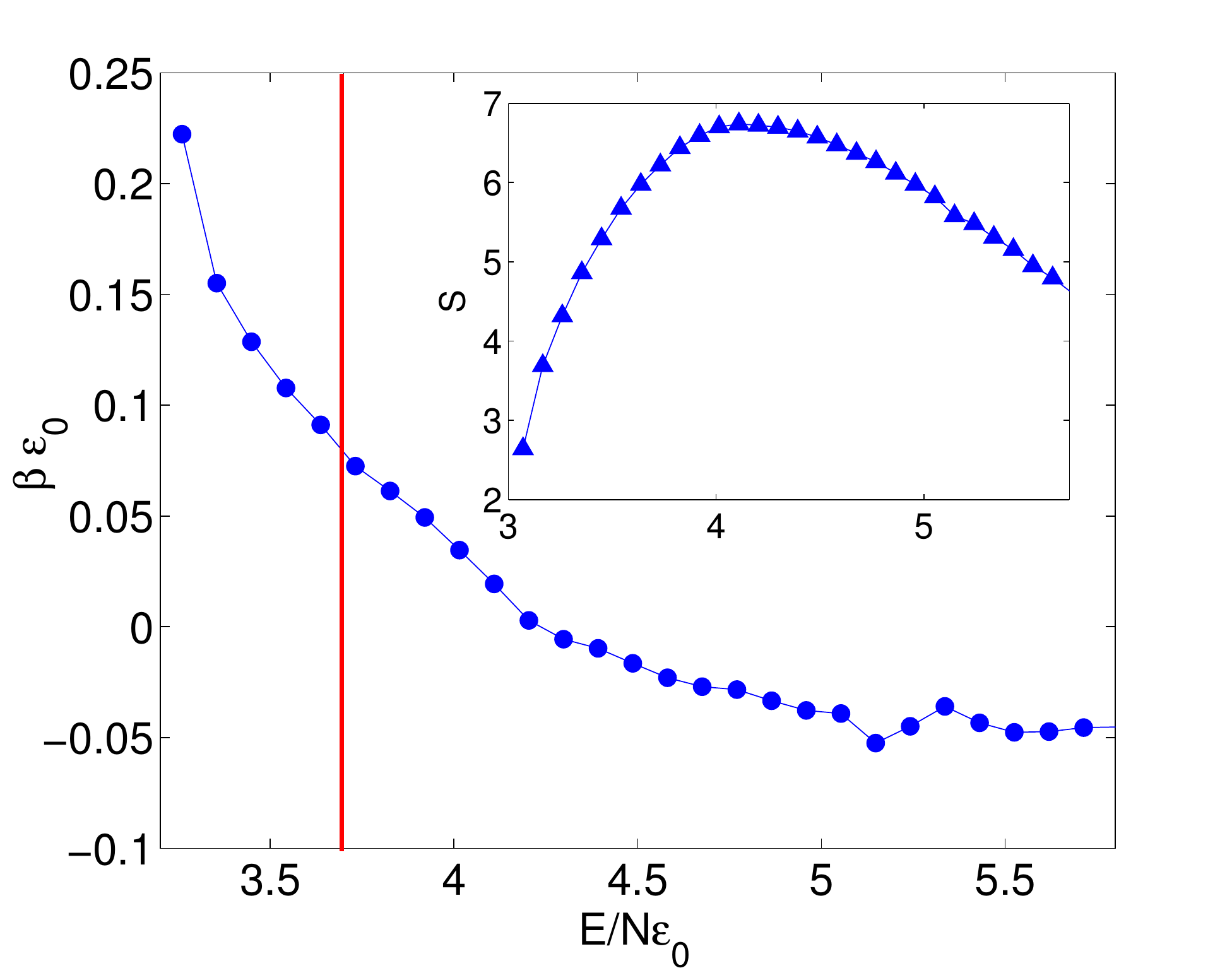}

\protect\caption{Inverse microcanonical temperature of the bath as a function of energy.
It can be inferred from the entropy showing in the inset using the
relation $\beta(E)=\partial S(E)/\partial E$. Dashed line shows energy
of the bath used in the calculations. \label{fig:T=000026S}}
\end{figure}

We study now the case $\hat{X}_{S}=\sigma_{z}$, which does not allow
energy exchange between the bath and the qubit. Following the same
procedure as above we arrive at the following master equations

\begin{eqnarray*}
\hbar\frac{\partial\rho_{12}}{\partial t} & = & -2g^{2}{\cal S}(0)\rho_{12}+i\Delta\rho_{12},\\
\hbar\frac{\partial\rho_{21}}{\partial t} & = & -2g^{2}{\cal S}(0)\rho_{21}-i\Delta\rho_{21},
\end{eqnarray*}
while $\partial\rho_{11}/\partial t=\partial\rho_{22}/\partial t=0$.
This leads to dephasing of the off-diagonal elements $|\rho_{12}(t)|=|\rho_{21}(t)|\sim\exp[-2g^{2}{\cal S}(0)t/\hbar]$
but not to thermalization, since $\rho_{11}$ and $\rho_{22}$ do
not change. 

These results are in line with the analysis in the paragraph below
Eq. (\ref{eq:Fluctuation2}): the first term in Eq. (\ref{eq:Fluctuations})
causes energy level shift of the qubit, while the residual thermal
fluctuations lead to the loss of coherence but not necessarily to
thermalization (cf. also Ref. \cite{fialko14}). Therefore, energy
exchange between the system and the bath is crucial to obtain thermalization
of the system. 

The Markovian assumption might not be strictly satisfied for a concrete
system. In this case, for example, the upper limit of the time integral
in Eq. (\ref{eq:MasterEQ}) cannot be extended to infinity but only
up to time $t$ \cite{breuer02}. This renders more fine-grained dynamics
especially at short times. As we are concerned with thermalization
of a qubit, we are interested in the asymptotic state at $t\rightarrow\infty$.
It is thus expected that the above formalism is adequate for the present
studies.

\section{Numerical Experiment}

To realize the proposed heat bath, we consider a two-band double-well
potential filled with cold bosons. A complex interplay between the
tunneling of bosons and their mutual interactions makes it possible
to satisfy the ETH criteria and show thermalization \cite{cosme14}.
We use this system as a heat bath for a qubit coupled to it. The Hamiltonian
of the bath reads \cite{cosme14,frazer07}

\begin{eqnarray}
\hat{H}_{B} & = & -\sum_{r\neq r',l}J^{l}\hat{b}_{r}^{l\dagger}\hat{b}_{r'}^{l}+\sum_{r,l}U^{l}\hat{n}_{r}^{l}(\hat{n}_{r}^{l}-1)+\sum_{r,l}E_{r}^{l}\hat{n}_{r}^{l}\nonumber \\
 & + & U^{01}\sum_{r,l\neq l'}(2\hat{n}_{r}^{l}\hat{n}_{r}^{l'}+\hat{b}_{r}^{l\dagger}\hat{b}_{r}^{l\dagger}\hat{b}_{r}^{l'}\hat{b}_{r}^{l'}),
\end{eqnarray}
where $\hat{b}_{r}^{l}$ ($\hat{b}_{r}^{l\dagger}$) are the annihilation
(creation) operators of a boson with mass $m_{B}$ in the well $r=L,R$
on the energy level $l=0,1$. For a given double well potential $V$(x)
the corresponding single-particle wave functions $\phi_{r}^{l}(x)$
can be found. The coefficients in the above equation can be then evaluated
\cite{frazer07} as follows: The single-particle tunneling rate is
$J^{l}=\int dx\phi_{L}^{l\ast}(x)\left(-\frac{\hbar^{2}}{2m_{B}}\nabla^{2}+V(x)\right)\phi_{R}^{l}(x)$,
on-site interaction strength is $U^{l}=g_{B}\int dx|\phi_{r}^{l}(x)|^{4}$,
single-particle energies are $E_{r}^{l}=\int dx\phi_{r}^{l\ast}(x)\left(-\frac{\hbar^{2}}{2m}\nabla^{2}+V(x)\right)\phi_{r}^{l}(x)$,
and induced interaction between levels is $U^{01}=g_{B}\int dx|\phi_{r}^{0}(x)|^{2}|\phi_{r}^{1}(x)|^{2}$.
Here, $g_{B}$ is the two-body interaction coupling between bosons. 

The double well is created by splitting the harmonic potential $m_{B}\omega_{0}^{2}x^{2}/2$
by the focused laser $10\hbar\omega_{0}\exp(-x^{2}/2\sigma^{2})$
with the width $\sigma=0.1\sqrt{\hbar/m_{B}\omega_{0}}$. We choose
energy units $\epsilon_{0}=\hbar\omega_{0}$. For $g_{B}=0.3\epsilon_{0}$
we obtain $J^{0}=0.26\epsilon_{0}$, $J^{1}=0.34\epsilon_{0}$, $U^{0}=0.14\epsilon_{0}$,
$U^{1}=0.1\epsilon_{0}$, $U^{01}=0.06\epsilon_{0}$, $E_{0}=1.25\epsilon_{0}$,
and $E_{1}=3.17\epsilon_{0}$. 

The Hamiltonian of the bath can be easily diagonalized numerically
for $N=30$ bosons and the eigenenergies $|E_{k}\rangle$ with the
corresponding eigenvalues $E_{k}$ can be extracted. The bath satisfies
the ETH criteria at low energies as it is shown in Fig. \ref{fig:ETH}.
We define the entropy of the bath at energy $E$ as $S(E)=-{\rm Tr}\left(\hat{\rho}_{m}{\rm ln}\hat{\rho}_{m}\right)$
and the corresponding inverse temperature as its derivative $\beta(E)=\partial S(E)/\partial E$.
The results are presented in Fig. \ref{fig:T=000026S}. 
\begin{figure}
\includegraphics[width=0.9\columnwidth]{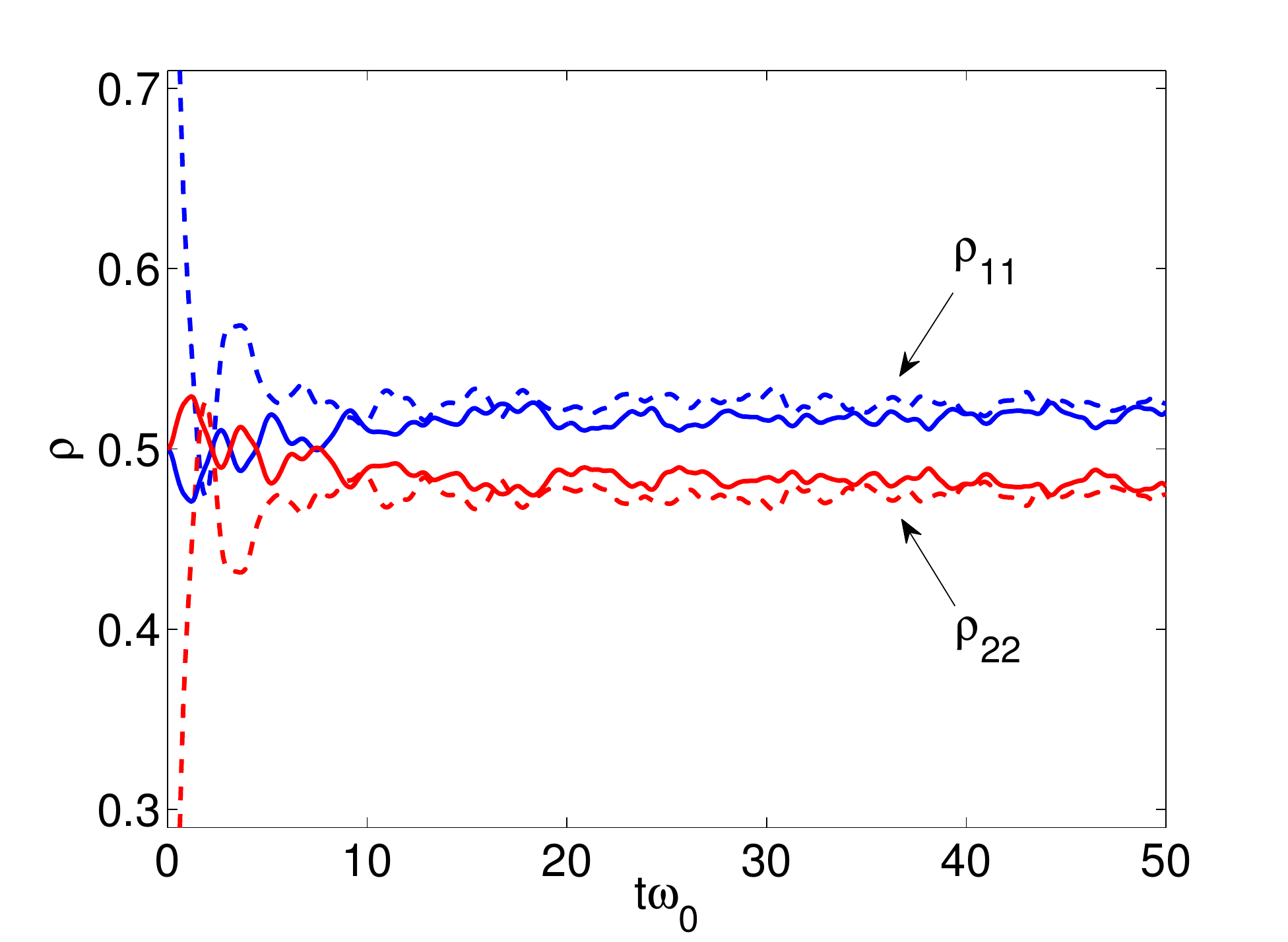}\protect\caption{Diagonal elements of the reduced density matrix of the qubit. It relaxes
to the thermal state $\rho_{11}/\rho_{22}=\exp(\beta\Delta)$. Solid
lines correspond to the initial state of the qubit $(1,1)^{T}/\sqrt{2}$,
dashed lines to $(1,0)^{T}$ respectively. For our simulations we
chose $\Delta=1\epsilon_{0}$ and we found $\beta\approx0.8\epsilon_{0}^{-1}$.
This is in very good agreement with the inverse microcanonical temperatue
of the bath; cf. Fig. \ref{fig:T=000026S}. Inset: corresponding energies
of the bath. \label{fig:Dynamics}}
\end{figure}

A two level qubit is represented via a single particle with mass $m_{S}$
trapped in a harmonic potential $m_{S}\omega^{2}x^{2}/2$ with two
energy states $\psi_{n}(x)$ ($n=0,1$) available to it. The corresponding
single-particle Hamiltonian is

\begin{equation}
\hat{H}_{S}=\hbar\omega\sum_{n=0}^{1}(n+1/2)\hat{a}_{n}^{\dagger}\hat{a}_{n}.
\end{equation}
 We choose $\Delta\equiv\hbar\omega=1\epsilon_{0}$. The particle
interacts with the bosons in the double well potential via contact
interaction $g$. The Hamiltonian of the combined system is given
in Eq. (\ref{eq:Hfull}), where the last term describes the interaction
between the qubit and the bath:
\begin{figure}
\includegraphics[width=0.9\columnwidth]{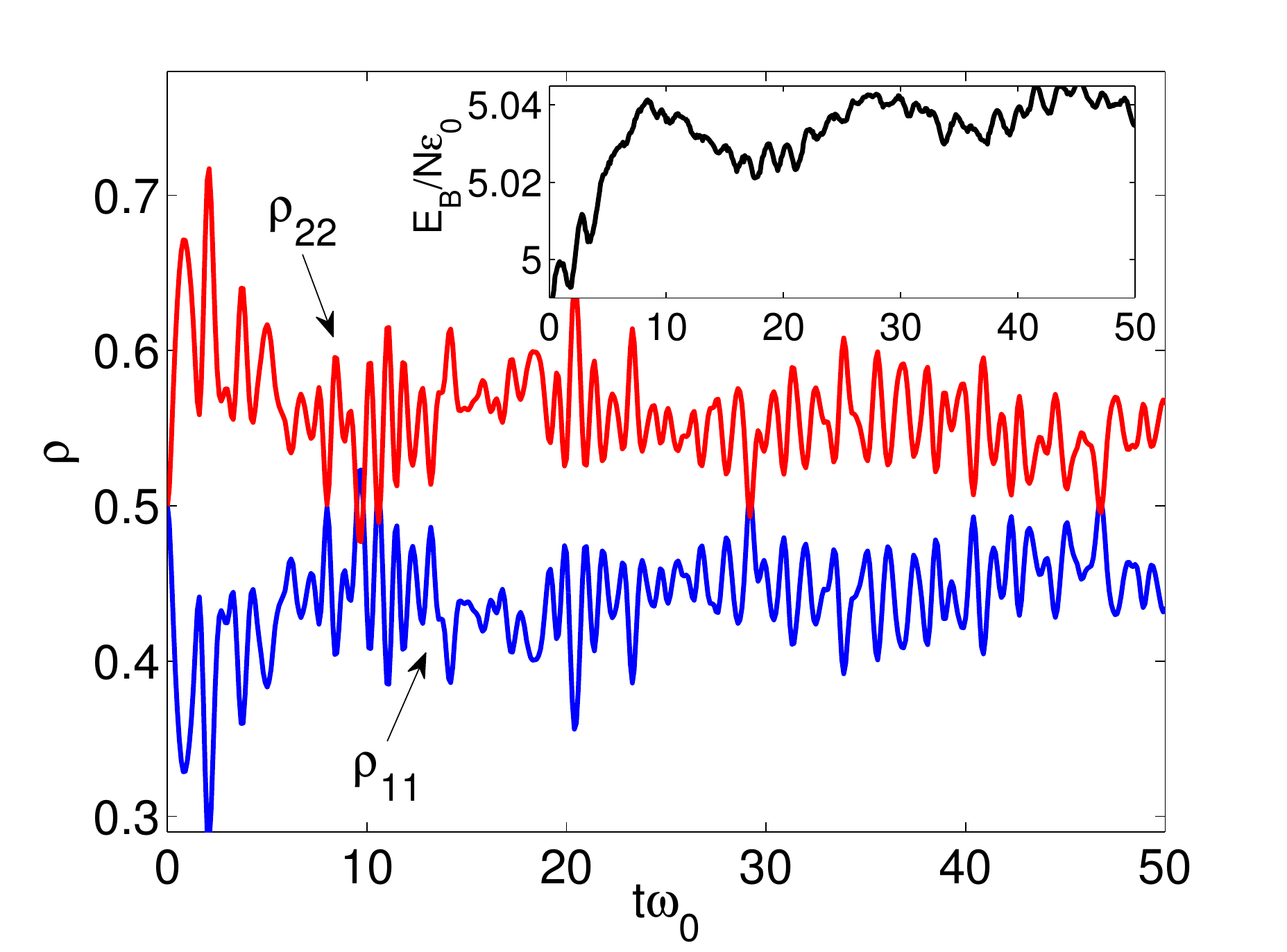}\protect\caption{The diagonal elements of the reduced density matrix of the qubit .
The initial state of the qubit is $(1,1)^{T}/\sqrt{2}$ and the initial
value of the bath energy $E_{B}\approx5N\epsilon_{0}$. ETH is not
satisfied at this energy, cf. Fig. \ref{fig:ETH}. The qubit does
not relax to the thermal state. Inset: corresponding energy of the
bath. \label{fig:Dynamics2}}
\end{figure}

\begin{equation}
\hat{H}_{I}=g\sum_{nn'll'=0}^{1}\sum_{r,r'=L}^{R}C_{rr'}^{nn'll'}\hat{a}_{n}^{\dagger}\hat{a}_{n'}\hat{b}_{r}^{l\dagger}\hat{b}_{r'}^{l'},
\end{equation}
where $C_{rr'}^{nn'll'}=\int dx\psi_{n}^{\ast}(x)\psi_{n'}(x)\phi_{r}^{l\ast}(x)\phi_{r'}^{l'}(x)$.
The inter level transitions of the qubit between $n\ne n'$ are allowed,
such that the qubit and the bath may exchange energy. We expect the
qubit to relax to a thermal state with the microcanonical temperature
of the bath. 

We simulate the quantum dynamics by creating the Hamiltonian in the
Fock basis of localized wavefunctions and propagate an initial state
$|\psi(0)\rangle$ by solving the Schödinger equation, $|\psi(t)\rangle=\exp(-iHt/\hbar)|\psi(0)\rangle$.
The initial state of the entire system is a Fock state of the bath
$|n_{L}^{0},n_{R}^{0},n_{L}^{1},n_{R}^{1}\rangle$ times an initial
state of the qubit. We choose two initial states for the qubit: $\left(1,0\right)^{T}$
corresponding to the particle occupying initially the lowest state
of the trapping potential and $\left(1,1\right)^{T}/\sqrt{2}$ corresponding
to the superposition of the lowest and the first excited states. The
initial Fock state of the bath is chosen such that the energy of the
bath $\langle\psi(0)|\hat{H}_{B}|\psi(0)\rangle\approx3.65N\epsilon_{0}$
satisfies the ETH. The energy of the bath $\langle\psi(t)|\hat{H}_{B}|\psi(t)\rangle$
changes in time slightly from this value, since the qubit is coupled
weakly to the bath and its Hilbert space is much smaller than that
of the bath. At this energy the inverse temperature of the bath can
be found from Fig. \ref{fig:T=000026S}, $\beta\approx0.8\epsilon_{0}^{-1}$.
As expected, the reduced density matrix of the qubit $\hat{\rho}(t)={\rm Tr}_{B}|\psi(t)\rangle\langle\psi(t)|$
relaxes to the thermal state $\lim_{t\rightarrow\infty}\rho_{11}(t)/\rho_{22}(t)=\exp(\beta\Delta)$
as shown in Fig. \ref{fig:Dynamics}. On the contrary, thermalization
is not observed for higher energies of the bath where ETH is not satisfied
as it is evident from Fig. \ref{fig:Dynamics2}. Therefore, ETH is
crucial to obtain thermalization of a small quantum system.

\section{Conclusions}

We have demonstrated that isolated quantum systems satisfying the
criteria of ETH can serve as finite autonomous heat baths for smaller
quantum systems. As an example, we studied theoretically and numerically
the thermal relaxation of a qubit weakly coupled to a realistic ETH
heat bath. We believe ETH heat baths will be of benefit in building
thermal machines working genuinely on the microscopic level, where
not only the engine but also heat bath is treated quantum mechanically. 

The qubit is believed to be the smallest quantum engine. It is an
interesting open question of how large the Hilbert space of the thermalized
system can get. 

The author acknowledges support from the Marsden Fund (Project No.
MAU1205), administrated by the Royal Society of New Zealand.

\end{document}